\documentclass[aps,prb,twocolumn,superscriptaddress]{revtex4}

\usepackage{graphicx}

\begin{document}

\title{Propagation of Avalanches in
Mn$_{12}$-acetate: Magnetic Deflagration }

\author{Yoko Suzuki}

\affiliation{Department of Physics, City College of New York, CUNY, New
York, NY 10031, USA}

\author{M. P. Sarachik}

\affiliation{Department of Physics, City College of New York, CUNY, New
York, NY 10031, USA}

\author{E. M. Chudnovsky}

\affiliation{Department of Physics and Astronomy, Lehman College, CUNY,
Bronx, 10468-1589, NY, USA}

\author{S. McHugh}

\affiliation{Department of Physics, City College of New York, CUNY, New
York, NY 10031, USA}

\author{R. Gonzalez-Rubio}

\affiliation{Department of Physics, City College of New York, CUNY, New
York, NY 10031, USA}

\author{Nurit Avraham}

\affiliation{Department of Condensed Matter Physics, the Weizmann
Institute of Science, Rehovot 76100, Israel}

\author{Y. Myasoedov}

\affiliation{Department of Condensed Matter Physics, the Weizmann
Institute of Science, Rehovot 76100, Israel}

\author{E. Zeldov}

\affiliation{Department of Condensed Matter Physics, the Weizmann
Institute of Science, Rehovot 76100, Israel}

\author{H. Shtrikman}

\affiliation{Department of Condensed Matter Physics, the Weizmann
Institute of Science, Rehovot 76100, Israel}

\author{N. E. Chakov}

\affiliation{Department of Chemistry, University of Florida,
Gainesville, FL 32611, USA}

\author{G. Christou}

\affiliation{Department of Chemistry, University of Florida,
Gainesville, FL 32611, USA}

\date{\today}

\begin{abstract}

Local time-resolved measurements of fast reversal of the
magnetization of single crystals of Mn$_{12}$-acetate indicate
that the magnetization avalanche spreads as a narrow interface
that propagates through the crystal at a constant velocity that is
roughly two orders of magnitude smaller than the speed of sound.
We argue that this phenomenon is closely analogous to the
propagation of a flame front (deflagration) through a flammable
chemical substance.

\end{abstract}


\maketitle

Mn$_{12}$-acetate (hereafter Mn$_{12}$-ac) is a prototypical
molecular magnet composed of magnetic molecules,

[Mn$_{12}$O$_{12}$(CH$_3$COO)$_{16}$(H$_2$O)$_4$]$\cdot$2CH$_3$COOH$\cdot$4H$_2$O,

\noindent with cores consisting of twelve Mn atoms strongly coupled by
exchange to form superparamagnetic clusters of spin $S = 10$ at low
temperatures \cite{Sessoli}.  Arranged in a centered tetragonal lattice,
the spin of the Mn$_{12}$ clusters is subject to strong magnetic
anisotropy along the symmetry axis (the c-axis of the crystal).  Below
the blocking temperature of $\approx 3.5\,$K, the crystal exhibits
remarkable staircase magnetic hysteresis due to resonant quantum spin
tunneling between energy levels on opposite sides of the anisotropy
barrier corresponding to different spin projections, as illustrated in
Fig. 1(a).\cite{Friedman}  This and other interesting properties of
Mn$_{12}$-ac have been intensively studied in the last decade (see Refs.
\onlinecite{review1,review2,friedmanreview,Barco} for reviews).

It has been known for some time \cite{Paulsen} that Mn$_{12}$-ac crystals
exhibit an abrupt reversal of their magnetic moment under certain
conditions.  This phenomenon, also observed in other molecular magnets,
has been attributed to a thermal runaway (avalanche) in which the initial
relaxation of the magnetization toward the direction of the field results
in the release of heat that further accelerates the magnetic relaxation.
Direct measurements of the heat emitted by Mn$_{12}$-ac
crystals,\cite{Fominaya} as well as measurements of the magnetization
reversal in pulsed magnetic fields,\cite{avalanches} have confirmed the
thermal nature of the avalanches.  More recently, the
electromagnetic signal associated with avalanches was detected
\cite{Tejada,Tejada2} and it was argued that if the radiation is of
thermal origin it would indicate a significant increase in the
temperature of the crystal.  This has not been confirmed by direct bulk
measurements of the temperature using a thermometer. Evidence has been
obtained \cite{Bal} that the avalanche may not be a uniform process
throughout the sample.  No clear understanding of the avalanche process
has emerged to date.

In this Letter we report local time-resolved measurements of fast
magnetization reversal (avalanches) in mm-size single crystals of
Mn$_{12}$-ac.  We show that a magnetic avalanche takes the form of a thin
interface between regions of opposite magnetization which propagates
throughout the crystal with a constant field-dependent speed ranging from
$1$ to $15$ m/s.  We demonstrate that this phenomenon is closely
analogous to the propagation of a flame front (deflagration) through a
flammable chemical substance.

Microscopic arrays of Hall bars were used to measure the magnetization of
three single crystals of Mn$_{12}$-ac with dimensions:
sample 1, $0.29 \times 0.29 \times 0.64$ mm$^3$; sample 2, $0.28 \times
0.28 \times 1.44$ mm$^3$; sample 3, $0.24 \times 0.24 \times 1.02$
mm$^3$.  Eleven Hall bars of dimensions $10 \times 10$ $\mu$m$^2$ with
$30$ $\mu$m intervals were used for sample 1, and $30 \times 30$
$\mu $m$^2$ with
$130$ $\mu$m intervals for samples 2 and 3.

\begin{figure}
\includegraphics[width=0.4\textwidth]{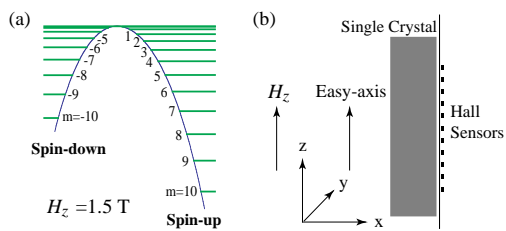}
\caption{\label{Fig1} (a)  The potential energy function in a
longitudinal magnetic field; (b) Schematic diagram of a sample mounted on
an array of Hall sensors used to detect $B_x$.  The magnetization is
relatively uniform for quasi-equilibrium conditions; pronounced
non-uniformity in magnetization during an avalanche generates
large values of $B_x$.}
\end{figure}

Using an excitation current of $2$ $\mu$A, the Hall bar signal was
amplified by a factor of $1200$, and detected and recorded by
several digital scopes and a data acquisition card.  In order to
ensure proper synchronization, one channel of each scope was
anchored to the same signal.  The Hall sensor and amplifier
introduced combined delays of up to $3$ $\mu$s.  A magnetic
field was applied in the z-direction along the crystal easy axis
(see Fig. 1), lowering (raising) the energy of the states corresponding
to spin projections along (opposite to) the field direction.  The Hall
sensors were aligned to detect the magnetic induction of the sample in
the $x$-direction.  $B_x$ is proportional to the spatial
derivative of $M_z$ in the  region near the sensor.  For a uniform
magnetization in the z-direction, $B_x$ derives from the gradient at
the  sample ends, which is proportional to $M_z$ itself.  During an
avalanche, there is a large contribution to $B_x$ from the local region
corresponding to the avalanche front, where $\partial M_z/\partial z$ is
large.

The samples were immersed in liquid $^3$He; most of the data were
obtained at the base temperature of $250$ mK.  The few points measured at
$400$ and $650$ mK were found to lie on the same curve within the scatter
of the data, indicating that the temperature dependence is weak.  A
longitudinal magnetic field (parallel to the easy axis) was swept back
and forth through the hysteresis loop to $\pm 6$ T until an avalanche
was triggered.  As reported in an earlier paper \cite{MMMsuzuki},
avalanches occur in a stochastic way at $0.25$ K both at resonant
magnetic fields (where energy levels on opposite side of the barrier
match, see Fig. 1) and away from resonance. Avalanches were also found
for sample 2 for zero field-cooled conditions, where the
sample starts from zero magnetization (instead of full
saturation).

\begin{figure}
\includegraphics[width=0.4\textwidth]{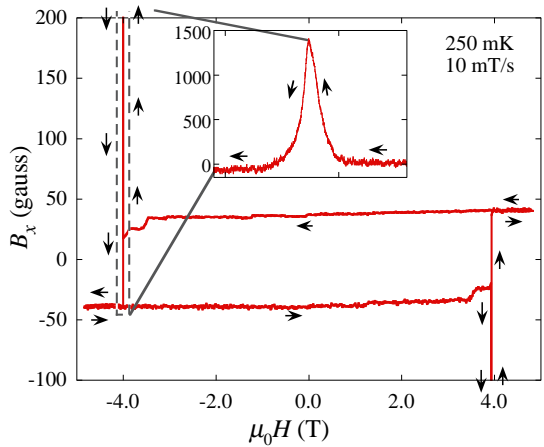}
\caption{\label{Fig2} Hysteresis loops for a single crystal of
Mn$_{12}$-ac, and an avalanche; data shown for sample 1.  $B_x$ was
recorded by a Hall sensor situated about $80$ $\mu$m above the middle of
the sample.  Steps due to quantum tunneling of the magnetization are
observed, followed by a sharp spike in $B_x$ associated with an avalanche,
shown in the inset. }
\end{figure}

Figure 2 shows an avalanche for sample 1.  Steps due to quantum tunneling
of the magnetization were observed, with a magnetization that was almost
uniform throughout the sample, until an avalanche occured, as shown in the
inset.  During the avalanche the Hall bar recorded a large peak in $B_x$,
signaling the abrupt onset of a highly non-uniform magnetization.

\begin{figure}
\includegraphics[width=0.4\textwidth]{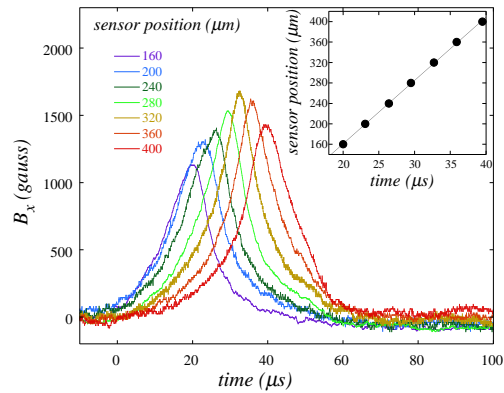}
\caption{\label{Fig3} Signals recorded by seven equally spaced Hall
sensors situated near the center of sample 1 during an avalanche
triggered at $4$ T.  The left (right) trace corresponds to the top
(bottom) sensor (see Fig. 1).  Sensor positions are measured relative
to the top of the sample, with the center at $\approx 320$ $\mu$m.  The
inset shows the sensor position versus the time at
which the sensor recorded peak amplitude.  A straight line fit yields a
constant velocity of propagation of $12$ m/s.}
\end{figure}

For a field sweep rate of $10$ mT/s and temperature $0.25$ K, Fig. 3 shows
an avalanche triggered at $4$ T and recorded for sample 1 by
seven of the eleven sensors placed in sequential positions near the center
of the sample.  The avalanche was triggered above the top-most sensor and
traveled downward (see Fig. 1).  $B_x$ displays the largest peak
at the center due to the finite size of the sample.  The inset shows the
sensor position as a function of the time at which the sensor registered
the peak amplitude.  The slope of the straight line drawn through these
points yields a constant velocity of $12$ m/s for this avalanche.

\begin{figure}
\includegraphics[width=0.4\textwidth]{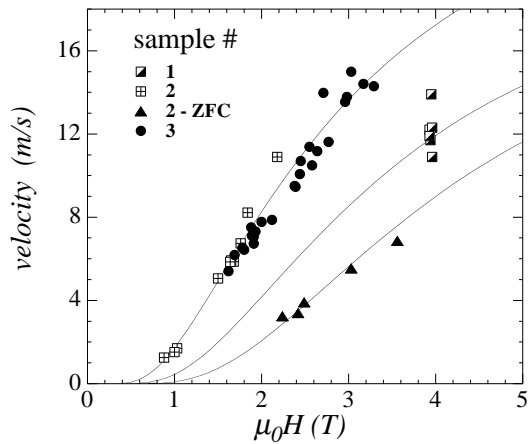}
\caption{\label{Fig4} Velocity of propagation of avalanches
versus longitudinal magnetic field at which the avalanches occured.
ZFC denotes data obtained for sample 2 cooled in zero field, thus
starting from zero magnetization and ending in full magnetization,
$\Delta M/2M_{sat} = 0.5$; $\Delta M/2M_{sat} = 1$ for sample 3 and
the remaining sample 2 data; for sample 4, $\Delta M/2M_{sat}$ varied
around $0.7$.  From top to bottom, the curves are fits to the data for
$\Delta M/2M_{sat} = 1, 0.7$ and $0.5$ (see text).}
\end{figure}

Figure 4 summarizes the data obtained for the velocity of propagation of
avalanches recorded in different longitudinal magnetic fields for all
three samples.  For avalanches starting from full magnetization, the data
for samples 2 and 3 lie on approximately the same curve.  The velocity
decreases with decreasing longitudinal magnetic field and goes to zero at
about $0.6$ T, below which no avalanches can occur.  Smaller velocities
are obtained for avalanches in sample 2 when starting from the
zero-field-cooled condition.  Avalanches for sample 1 were obtained only
at relatively high magnetic fields in the vicinity of $4$ T; the
velocities for this sample range in value and do not appear
to be consistent with data for the other two samples.

Interestingly, as shown in Fig. 5, an approximate collapse is obtained for
all the data when plotted as a function of $g\mu_B H S (\Delta
M/M_{sat})$, the energy per molecule released during an avalanche. 
Thus, avalanches require the release of a threshold energy, above which
they propagate with a speed that appears to be a linear function of the
energy for the range investigated in these experiments.

\begin{figure}
\includegraphics[width=0.4\textwidth]{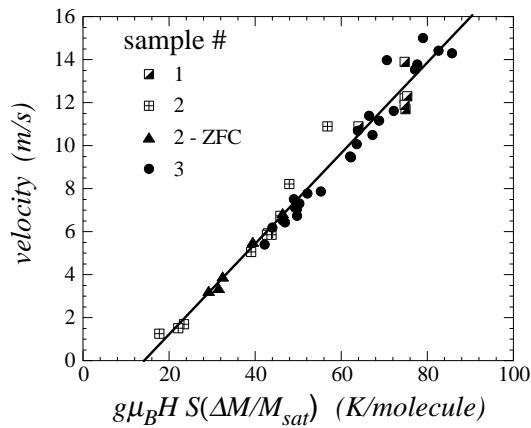}
\caption{\label{Fig5} Velocity of propagation of avalanches
versus energy released per molecule.  The line is
drawn to guide the eye.}
\end{figure}

Our observations cannot be attributed to magnetization reversal
associated with domain wall motion, since there is no long-range order in
our system.  Some insight can be obtained by noting that,
from a thermodynamic point of view, a crystal of Mn$_{12}$ molecules
placed in a magnetic field opposite to the magnetic moment is equivalent
to a metastable (flammable) chemical  substance.  In our case, the role
of the chemical energy stored in a molecule is played by the difference
in the Zeeman energy, $\Delta E = 2g\mu_BHS$, for states of the
Mn$_{12}$-ac molecule that correspond to ${\bf S}$ parallel and
antiparallel to ${\bf H}$; here $g = 1.94$ is the gyromagnetic factor and
$\mu_B$ is the Bohr magneton.  For Mn$_{12}$-ac in a field of a few Tesla,
$\Delta E$ is below $0.01\,$eV, as is the energy barrier, $U(H)$,
between spin-up and spin-down states due to the magnetic
anisotropy.  Thus, for the avalanches in Mn$_{12}$-ac, $\Delta E$
and $U$ are two orders of magnitude smaller than typical energies
of chemical reactions.  However, our temperature range is also
more than two orders of magnitude below room temperature,
making the analogy rather close.

A well-known mechanism for the release of energy by a metastable
chemical substance is combustion or slow burning, technically
referred to as deflagration.\cite{LL}  It occurs as a flame of
finite width, $\delta$, propagates at a constant speed, $v$, small
compared to the speed of sound.  The parameter $\delta$ is
determined by the distance, $\delta \sim \sqrt{\kappa \tau}\,$
through which the heat diffuses during the time of the ``chemical
reaction''
$\tau$. In our case
\begin{equation}\label{tau}
\tau(H) = \tau_0 \exp{\left[\frac{U(H)}{k_BT_f}\right]}\,,
\end{equation}
where $\tau_0 \sim 10^{-7}\,$s is the attempt time \cite{Friedman}
and $T_f$ is the temperature of the flame.  The dynamics of the
flame are governed by the thermal diffusivity, $\kappa$, which obeys:
\begin{equation}\label{conductivity}
\frac{\partial T}{\partial t} = \kappa \nabla^2T\,.
\end{equation}
For $\kappa$ independent of $T$, substituting $T = T(x - vt)$ at
$x > vt$, one obtains $T = T_f \exp{[-v(x - vt)/\kappa]}$ in front
of the interface, which yields $v \delta = \kappa$.  An interface
thickness that is at most the distance between sensors, $\delta \sim 30
\mu$m, and the experimentally measured velocities of
the order of $1 - 15$ m/s,  yield an upper bound on $\kappa$ in the range 
$10^{-5}\,$m$^2$/s to $10^{-4}\,$m$^2$/s, consistent with heat pulse
experiments.\cite{javier}

Combining $v \delta = \kappa$ with $\delta \sim \sqrt{\kappa \tau}\,$,
one obtains:
\begin{equation}\label{velocity}
v \sim \frac{\delta}{\tau} \sim \sqrt{\kappa/\tau} =
\left(\frac{\kappa}{\tau_0}\right)^{1/2}\exp\left[
-\frac{U(H)}{2k_BT_f}\right]\,.
\end{equation}
The strongest dependence of $v$ on $H$ derives from the exponential,
which contains the known dependence\cite{friedman2} of the energy
barrier, $U(H)$, on the magnetic field.  Assuming that the temperature of
the flame is proportional to the released magnetic energy density,
$T_f=CH\Delta M$, it is possible to fit all the data with a single
value of the proportionality constant $C$, as shown in Fig. 4. 
The flame temperature obtained from these fits ranges from $8.5$ K for an
avalanche at $1$ T to $26$ K for an avalanche triggered at $4$ T.  We
note that the prefactor obtained from the fit is consistent with the
values of $\kappa$ and $\tau_0$ discussed earlier.\cite{note}  Here we
have used the simplest model of deflagration, a widely studied phenomenon
that is known to be quite complex.\cite{Combustion}  This crude model
captures the overall behavior, and yields parameters that are quite
reasonable in size.  A more complete theory is needed to account for the
apparent data collapse of Fig, 5.

Recent bolometer measurements of
the radiation generated by a magnetic avalanche \cite{Tejada,Tejada2} gave
puzzling results that can be understood within our model.  One enigma was
that the sample temperature measured by a thermometer directly
following an avalanche was lower ($< 6\,$K) than the temperature
registered by the bolometer if one assumed thermal radiation.  A second
puzzle was that the reversal of the magnetization during the avalanche
occured on a much shorter time scale than the cooling of the sample
following the avalanche.\cite{Tejada2}  We suggest that the radiation
observed during avalanches is generated by the narrow hot interface
(flame) that propagates through the crystal.  The temperature of the bulk
of the crystal (including the ``ash'' left behind the interface) is
always significantly lower than the temperature of the interface itself. 
The time of the magnetization reversal is determined by the time $t=l/v$
needed for the interface to sweep the sample of length $l$. In our case
$t \sim 0.1\,$ms, while the time needed for the ``ash'' to reach
equilibrium with the thermal bath can be much longer.

The strongest evidence that our observations are due to
deflagration is the presence of a well defined propagating front
requiring a threshold energy traveling at a subsonic
velocity.  The deflagration mechanism provides the condition needed for
the avalanche to occur.  This condition is the same as the condition
needed to sustain the propagation of a flame through a chemical
substance.  It is well-known that deflagration of a flammable gas will
not occur in a pipe of diameter,
$d$, small compared to the width of the flame, $\delta$.  In our case 
$\delta$ must be small compared to the diameter of the crystal. If this
condition is not satisfied, the heat generated by the magnetization
reversal diffuses mostly through the walls of the sample and cannot
sustain the propagation of the interface.  This explains why avalanches
only occur in larger crystals with sufficiently large magnetization
opposite to the direction of the field.  The latter condition coincides
with the condition of ``flammability" \cite{Combustion} needed to provide
sufficient heating (that is, the large $T_f$) required for
$\delta < d$.  It is interesting to note in this connection that the few
avalanches that were recorded around $1$ T did not result in full
reversal of the magnetization.  At these low fields the conditions for
ignition are marginally satisfied, and the ``flame'' is extinguished
before the process of magnetization reversal has been completed.  In
addition to available magnetic energy (flammability), the conditions for
ignition may also depend on the shape and quality of the  crystal, which
may account for differences observed for different samples.

Slow burning at a subsonic speed (deflagration) is governed by the
linear process of thermal conductivity.  In addition to deflagration,
unstable chemical substances also exhibit detonation, which can be
caused by instability of the flame or by direct initiation other than
through deflagration.\cite{Combustion}  The initial stage of the
detonation corresponds to a non-linear supersonic shock
wave.\cite{LL,Combustion}  Theory and experimental studies of advanced
stages of detonation are lacking.  Based on the close analogy between
unstable chemical substances and molecular magnets, the latter may well
exhibit ``magnetic detonation" under the right conditions.

In conclusion, we have demonstrated that avalanches in the
magnetization reversal of sufficiently large crystals of magnetic
molecules are very similar to flame propagation (deflagration)
through a metastable chemical substance.  The analogy between the
two systems derives from the magnetic bi-stability of molecular
nanomagnets.  Our observation of ``magnetic deflagration'' offers a
potentially important new way to investigate the phenomenon of flame
propagation (and, possibly detonation).  In contrast to deflagration in
flammable chemical substances, the analogous process of ``magnetic
deflagration'' in molecular nanomagnets is non-destructive, reversible,
and much easier to control.

We are grateful to D. Graybill for participation in various aspects of
this work, and to K. M. Mertes, J. R. Friedman, M. Bal, and J. Tejada for
valuable discussions.  Work at City College was supported by NSF grant
DMR-0451605.  Support was provided for EMC by NSF Grant
No. EIA-0310517, and for GC by NSF Grant No. CHE-0071334.  E. Z.
acknowledges the support of the Israeli Science Foundation Center of
Excellence.


\begin{thebibliography}{99}
\bibitem{note} The prefactor of the exponential in Eq. \ref{velocity}, the
second parameter required for the fits, differed by as much as $30$\% for
different $\Delta M$, reflecting the fact that the prefactor is not well
known within the theory.  On the other hand, the exponential dependence
is well captured by the model.
\end{thebibliography}
\end{document}